\documentclass{llncs}

\usepackage{amsmath,amsfonts,amssymb}
\usepackage{pst-all}
\usepackage{epsfig}
\usepackage{epsf}
\usepackage{here}
\usepackage{graphicx}
\usepackage{graphics}
\usepackage{subfigure}
\usepackage{slashbox}
\usepackage{multirow}
\usepackage{shadow}
\usepackage{psboxit}

\newcommand{\DC}{\ensuremath{C}}
\newcommand{\PT}[1]{\ensuremath{{\DC}_{#1}}}
\newcommand{\PTS}[2]{\ensuremath{{\DC}_{#1,#2}}}
\newcommand{\SPRED}[2]{\ensuremath{S(#1,#2)}}

\newcommand{\VEC}[1]{\ensuremath{{\bf #1}}}

\def\qed{\relax\ifmmode\hskip2em \Box\else\unskip\nobreak\hskip1em $\Box$\fi}
\newcommand{\refLemma}[1]{Lemma~\ref{#1}}

\newcommand{\refFigure}[1]{Fig.~\ref{#1}}
\newcommand{\refProposition}[1]{Proposition~\ref{#1}}
\newcommand{\refTheorem}[1]{Theorem~\ref{#1}}
\newcommand{\Equ}[1]{Eq.~(\ref{#1})}

\begin{document}

%
%
\pagestyle{headings}

\mainmatter

\title{Revisiting digital straight segment recognition}


\author{Fran\c{c}ois de Vieilleville \and Jacques-Olivier Lachaud }

\authorrunning{F. de Vieilleville \and J.-O. Lachaud}

\tocauthor{Fran\c{c}ois de Vieilleville, Jacques-Olivier Lachaud (LaBRI, Universit\'{e} Bordeaux 1)}

\institute{LaBRI, Univ. Bordeaux 1 \\ 351 cours de la Lib\'{e}ration,\\ 33405 Talence Cedex, France \\
\email{\{devieill,lachaud\}@labri.fr}}

\maketitle

\begin{abstract}
  This paper presents new results about digital straight segments,
  their recognition and related properties. They come from the study
  of the arithmetically based recognition algorithm proposed by
  I. Debled-Rennesson and J.-P. Reveill\`{e}s in 1995
  \cite{Debled95}. We indeed exhibit the relations describing the
  possible changes in the parameters of the digital straight segment
  under investigation.  This description is achieved by considering
  new parameters on digital segments: instead of their arithmetic
  description, we examine the parameters related to their combinatoric
  description. As a result we have a better understanding of their
  evolution during recognition and analytical formulas to compute
  them. We also show how this evolution can be projected onto the
  Stern-Brocot tree. These new relations have interesting consequences
  on the geometry of digital curves. We show how they can for instance
  be used to bound the slope difference between consecutive maximal
  segments.
\end{abstract}


\section{Introduction}
\label{sec:intro}

The study of digital straight lines is a fundamental topic in discrete
geometry and several approaches have been proposed. Following the
taxonomy of \cite{rKaR04a}, we can divide them into three groups.  The
first one characterizes digital lines through the study of the {\em
pre-image}: this aims at determining in a dual space the possible real
lines whose digitization corresponds to a given set of pixels
\cite{AS016}. That kind of approaches has recently been used to define
and recognize straight lines drawn on irregular isothetic grids
\cite{Coeurjolly05}.

A second group is related to combinatorics and relies on the link
between continued fractions and recursive characterization of digital
lines. It can cope with lines with rational or irrational slopes, as
their digitization can be seen as a word over a finite alphabet. The
tools developed to characterize and study those objects
\cite{Berstel97,Voss93} have for instance been used to study the
asymptotic behavior of some digital segments over digitizations of
$\mathcal{C}^{3}$ convex curves \cite{deVieilleville05a,devieill05}.


The third group gathers arithmetic approaches, which are based on a
formulation very similar to the one of real lines (Diophantine
inequalities, slope and vertical shift). They have led to simple,
incremental and optimal algorithms to recognize digital segments
\cite{Debled95,Feschet99}. For this approach the best known
recognition algorithm is the above mentioned algorithm of Debled and
Reveill\`{e}s \cite{Debled95}, referenced as {\bf DR95} algorithm in
the recent book of Klette and Rosenfeld \cite{Klette04}. This
algorithm extracts progressively the most simple digital line
parameters of a finite connected sequence of pixels, updating the
parameters at each pixel.

In this algorithm, the parameters refer to an arithmetic
representation of digital straight lines: slope as a fraction, integer
shift to origin, position of some specific limit points (upper and
lower leaning points). Their evolution during the progressive steps of
the recognition is governed by algorithmic computations: for instance,
the new slope is computed from the last added point and some former
leaning point. Although sufficient for recognizing digital lines,
these parameters lack of descriptive content to fully understand what
is digital straightness. For instance, they cannot answer a question
like if two straight lines share a common part, how are related their
slopes. Along the same lines, although it is known since Debled's
thesis \cite{Debled95t} that slope evolutions during recognition
correspond to displacements in the Stern-Brocot tree, these parameters
are nevertheless incomplete to fully describe it.


We propose here to use the combinatoric approach to give better
insights about the {\bf DR95} algorithm. A digital line is then
characterized by the continued fraction of its slope and the number of
patterns it contains. The evolution of these new parameters is then
precisely stated with analytic formulas. We also give another
interpretation of their evolution, as definite displacements on the
Stern-Brocot tree. Afterwards we focus on a particular class of
digital segments subset of digital curves, which are called
\emph{maximal segments} and which have interesting properties
\cite{Feschet03,Lachaud05a,Doerksen04}. Informally, they form the
inextensible digital straight segments on the curve. The preceding
properties allow us to give an analytic writing of the minimal and
maximal slope variation between two consecutive maximal
segments. These bounds are fully described with our new
parameters. Surprisingly, they show that consecutive maximal segments
may not vary too much nor too little since both bounds are of the same
order wrt parameters. On a long term, these quantitative relations
will be crucial for designing digital curvature estimators based on
slope variations.

This paper is organized as follows. First, basic arithmetic and
combinatoric definitions and properties of digital lines are recalled
(Section~\ref{sec:dss}). Then we give a comprehensive explanation of
the {\bf DR95} algorithm, describing each possible evolution in terms
of the new parameters (Section~\ref{sec:dr95}). Afterwards the
connection between the {\bf DR95} and the Stern-Brocot tree is
explicited (Section~\ref{sec:sbt}). Eventually those properties have
consequences on the geometry of maximal segments, namely bounds on
their slope variations (Section~\ref{sec:ms}). We conclude the paper
by some perspectives to this work (Section~\ref{sec:conclusion}).

\section{Digital straight segments: arithmetic and combinatoric approach}
\label{sec:dss}

Given a compact set with rectifiable boundary we consider its Gauss
digitization. The digital border of this digitization is chosen as the
inter-pixel 4-connected path laying between its inner and outer
digitization. This digital curve is referred as $\DC$.  We consider
that the points on the boundary are indexed increasingly, for instance
with a counterclockwise order. Moreover, given two points on the
boundary (say $A$ and $B$), $\PTS{A}{B}$ is the digital path from $A$
to $B$. For convenience reasons we identify the index of a point on
the boundary to the point itself.  For example $A < B$ means that the
point $A$ is before the point $B$ on the curve.

\begin{figure}[t] 
  \begin{center}
    \begin{picture}(0,0)%
\includegraphics{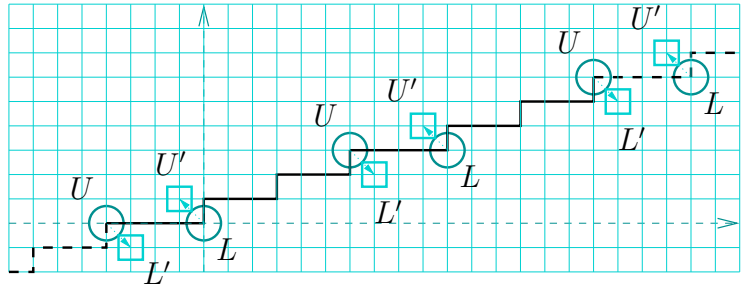}%
\end{picture}%
\setlength{\unitlength}{2013sp}%
\begingroup\makeatletter\ifx\SetFigFont\undefined%
\gdef\SetFigFont#1#2#3#4#5{%
  \reset@font\fontsize{#1}{#2pt}%
  \fontfamily{#3}\fontseries{#4}\fontshape{#5}%
  \selectfont}%
\fi\endgroup%
\begin{picture}(9167,3556)(2668,-3005)
\put(4351,-2911){\makebox(0,0)[lb]{\smash{{\SetFigFont{12}{14.4}{\familydefault}{\mddefault}{\updefault}{\color[rgb]{0,0,0}$L'$}%
}}}}
\put(5251,-2611){\makebox(0,0)[lb]{\smash{{\SetFigFont{12}{14.4}{\familydefault}{\mddefault}{\updefault}{\color[rgb]{0,0,0}$L$}%
}}}}
\put(8251,-1711){\makebox(0,0)[lb]{\smash{{\SetFigFont{12}{14.4}{\familydefault}{\mddefault}{\updefault}{\color[rgb]{0,0,0}$L$}%
}}}}
\put(11251,-811){\makebox(0,0)[lb]{\smash{{\SetFigFont{12}{14.4}{\familydefault}{\mddefault}{\updefault}{\color[rgb]{0,0,0}$L$}%
}}}}
\put(10351,239){\makebox(0,0)[lb]{\smash{{\SetFigFont{12}{14.4}{\familydefault}{\mddefault}{\updefault}{\color[rgb]{0,0,0}$U'$}%
}}}}
\put(7351,-661){\makebox(0,0)[lb]{\smash{{\SetFigFont{12}{14.4}{\familydefault}{\mddefault}{\updefault}{\color[rgb]{0,0,0}$U'$}%
}}}}
\put(4501,-1561){\makebox(0,0)[lb]{\smash{{\SetFigFont{12}{14.4}{\familydefault}{\mddefault}{\updefault}{\color[rgb]{0,0,0}$U'$}%
}}}}
\put(6451,-961){\makebox(0,0)[lb]{\smash{{\SetFigFont{12}{14.4}{\familydefault}{\mddefault}{\updefault}{\color[rgb]{0,0,0}$U$}%
}}}}
\put(9451,-61){\makebox(0,0)[lb]{\smash{{\SetFigFont{12}{14.4}{\familydefault}{\mddefault}{\updefault}{\color[rgb]{0,0,0}$U$}%
}}}}
\put(10201,-1261){\makebox(0,0)[lb]{\smash{{\SetFigFont{12}{14.4}{\familydefault}{\mddefault}{\updefault}{\color[rgb]{0,0,0}$L'$}%
}}}}
\put(7201,-2161){\makebox(0,0)[lb]{\smash{{\SetFigFont{12}{14.4}{\familydefault}{\mddefault}{\updefault}{\color[rgb]{0,0,0}$L'$}%
}}}}
\put(3451,-1861){\makebox(0,0)[lb]{\smash{{\SetFigFont{12}{14.4}{\familydefault}{\mddefault}{\updefault}{\color[rgb]{0,0,0}$U$}%
}}}}
\end{picture}%
  \end{center}
  \caption{Positions of weakly exterior points on a digital
    straight line of characteristics $(3,10,12)$. Weakly exterior
    points are boxed and leaning points are circled.
    \label{fig:pointOrder}}
\end{figure}

Let us recall the arithmetic definition of digital straight lines and
explain the notations that will be used in the following of the paper.
Following definitions hold in the first octant.
\begin{definition}
The set of points $(x,y)$ of the digital plane verifying $\mu \le
ax-by < \mu + |a|+|b|$, with $a$, $b$ and $\mu$ integer numbers, is
called the {\em standard line} with slope $a/b$ and shift $\mu$
\cite{Rev91} (e.g. see \refFigure{fig:pointOrder}).
\end{definition}
The {\em standard lines} are the 4-connected discrete lines. The
quantity $r_{(a,b)}(P)=ax-by$ is the {\em remainder} of the point
$P=(x,y)$ in the digital line of characteristics $(a,b,\mu)$. The
points whose remainder is $\mu$ (resp. $\mu+|a|+|b|-1$) are called
{\em upper} (resp.  {\em lower}) {\em leaning points}. 
\begin{definition} 
  A set of successive points $\PTS{i}{j}$ of the digital curve $\DC$
  is a {\em digital straight segment (DSS)} iff there exists a
  standard line $(a,b,\mu)$ containing them. The predicate
  ``$\PTS{i}{j}$ is a DSS'' is denoted by $\SPRED{i}{j}$. When
  $\SPRED{i}{j}$ the {\em characteristics} associated with the DSS
  $\PTS{i}{j}$ (extracted with the {\bf DR95} algorithm)
  \cite{Debled95} are the characteristics $(a,b,\mu)$ which minimize
  $a+b$. 
\end{definition}

The original {\bf DR95} algorithm recognizes naive digital straight
line but it is easily adapted to standard lines. It extracts the
characteristics $(a,b,\mu)$, with minimal $a+b$. The evolution of the
characteristics is based on a simple test: each time we try to add a
new point 4-connected to the current digital straight segment, we
compute its \emph{remainder} with respect to the DSS parameters.
According to this value the point can be added or not. If it is
greater than or equal to $\mu + a + b +1$ or less than or equal to
$\mu -2$ the point is said to be \emph{exterior} to the digital
straight segment and cannot be added. Otherwise the point can be added
to the segment to form a longer DSS and falls into two categories:
\begin{itemize} 
\item {\em interior} points, with a remainder between $\mu$ and $\mu
+a + b -1$ both included;
\item {\em weakly exterior} points, with a remainder of $\mu -1$ for
  {\em upper weakly exterior} points and $\mu + a + b$ for {\em lower
  weakly exterior} points. Only in this case are the characteristics updated.
\end{itemize}

Even if the arithmetic approach is a powerful tool for digital
straight segment recognition, other approaches may reveal useful to
get analytic properties. We here recall one of those approaches which
is connected to continued fractions.

\begin{definition}
  Given a standard line of characteristics $(a,b,\mu)$, we call
  \emph{pattern} of characteristics $(a,b)$ the word formed by the
  Freeman codes between any two consecutive upper leaning points. The
  Freeman codes defined between any two consecutive lower leaning
  points is the previous word read from back to front and is called
  the \emph{reversed pattern} of characteristics $(a,b)$.
\end{definition}

Since a DSS has at least either two upper or two lower leaning points,
a DSS $(a,b,\mu)$ contains at least one pattern \emph{or} one reversed
pattern of characteristics $(a,b)$. It is important to note that a DSS
$(a,b,\mu)$ contains $\delta$ pattern $(a,b)$ (resp.  $\delta'$
reversed-pattern $(a,b)$) iff it has $\delta +1$ upper leaning points
(resp. $\delta' +1$ lower leaning points). Moreover for any
DSS$(a,b,\mu)$, the number of pattern $(a,b)$ and reversed-pattern
$(a,b)$ differ from one.


There exists recursive transformations for computing the
\emph{pattern} of a standard line from the \emph{simple continued
  fraction} of its slope (\cite{Berstel97}, \cite{Klette04} Chap.~9
or \cite{Voss93} Chap.~4 ). We chose to focus on Berstel's approach,
which better suits our purpose.
A \emph{continued fraction} $z$ will be conveniently denoted by
$[0,u_{1} \ldots, u_n, \ldots]$.  The $u_{i}$ are called
\emph{elements} or \emph{partial coefficients} and the continued
fraction formed with the $k+1$ first \emph{partial coefficients} of
$z$ is said to be a \emph{$k$-th convergent} of $z$ and is denoted
$z_{k}$. The \emph{depth} of a $k$-th convergent equals $k$. We
conveniently denote by $p_{k}$ the numerator and by $q_{k}$ the
denominator of a $k$-th convergent.

We recall a few more relations regarding the way convergents can be
formed: 
\begin{eqnarray}
&  \forall k \geq 1 & \quad p_{k}q_{k-1} - p_{k-1}q_{k} = (-1)^{k+1}, \label{pattern:rec:dif} \\
p_{0} = 0 \quad p_{-1} = 1 \quad & \forall k \geq 1 & \quad   p_{k}=u_{k}p_{k-1}+p_{k-2}, \label{pattern:rec:num} \\
q_{0} = 1 \quad q_{-1}= 0  \quad & \forall k \geq 1 & \quad  q_{k}=u_{k}q_{k-1}+q_{k-2}. \label{pattern:rec:den}  
\end{eqnarray} 

Continued fractions can be finite or infinite, we focus on the case of
rational slopes of lines in the first octant, that is finite continued
fractions between $0$ and $1$. For each $i$, $u_{i}$ is assumed to be
a strictly positive integer. In order to have a unique writing we
consider that the last \emph{partial coefficient} is greater or equal
to two; except for slope $1 = [0,1]$.

\begin{figure}[t] 
    \begin{center}
\begin{picture}(0,0)%
\includegraphics{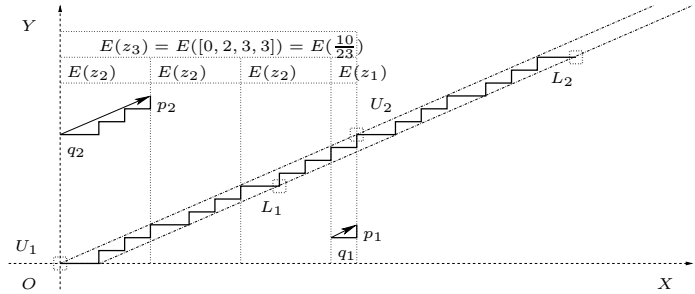}%
\end{picture}%
\setlength{\unitlength}{1066sp}%
\begingroup\makeatletter\ifx\SetFigFont\undefined%
\gdef\SetFigFont#1#2#3#4#5{%
  \reset@font\fontsize{#1}{#2pt}%
  \fontfamily{#3}\fontseries{#4}\fontshape{#5}%
  \selectfont}%
\fi\endgroup%
\begin{picture}(15924,6706)(1789,-7055)
\put(2101,-961){\makebox(0,0)[lb]{\smash{{\SetFigFont{6}{7.2}{\familydefault}{\mddefault}{\updefault}{\color[rgb]{0,0,0}$Y$}%
}}}}
\put(16876,-6961){\makebox(0,0)[lb]{\smash{{\SetFigFont{6}{7.2}{\familydefault}{\mddefault}{\updefault}{\color[rgb]{0,0,0}$X$}%
}}}}
\put(7651,-5161){\makebox(0,0)[lb]{\smash{{\SetFigFont{6}{7.2}{\familydefault}{\mddefault}{\updefault}{\color[rgb]{0,0,0}$L_{1}$}%
}}}}
\put(14401,-2161){\makebox(0,0)[lb]{\smash{{\SetFigFont{6}{7.2}{\familydefault}{\mddefault}{\updefault}{\color[rgb]{0,0,0}$L_{2}$}%
}}}}
\put(10201,-2761){\makebox(0,0)[lb]{\smash{{\SetFigFont{6}{7.2}{\familydefault}{\mddefault}{\updefault}{\color[rgb]{0,0,0}$U_{2}$}%
}}}}
\put(3151,-2011){\makebox(0,0)[lb]{\smash{{\SetFigFont{6}{7.2}{\familydefault}{\mddefault}{\updefault}{\color[rgb]{0,0,0}$E(z_{2})$}%
}}}}
\put(5251,-2011){\makebox(0,0)[lb]{\smash{{\SetFigFont{6}{7.2}{\familydefault}{\mddefault}{\updefault}{\color[rgb]{0,0,0}$E(z_{2})$}%
}}}}
\put(7351,-2011){\makebox(0,0)[lb]{\smash{{\SetFigFont{6}{7.2}{\familydefault}{\mddefault}{\updefault}{\color[rgb]{0,0,0}$E(z_{2})$}%
}}}}
\put(9451,-2011){\makebox(0,0)[lb]{\smash{{\SetFigFont{6}{7.2}{\familydefault}{\mddefault}{\updefault}{\color[rgb]{0,0,0}$E(z_{1})$}%
}}}}
\put(5251,-2761){\makebox(0,0)[lb]{\smash{{\SetFigFont{6}{7.2}{\familydefault}{\mddefault}{\updefault}{\color[rgb]{0,0,0}$p_{2}$}%
}}}}
\put(10051,-5761){\makebox(0,0)[lb]{\smash{{\SetFigFont{6}{7.2}{\familydefault}{\mddefault}{\updefault}{\color[rgb]{0,0,0}$p_{1}$}%
}}}}
\put(3151,-3811){\makebox(0,0)[lb]{\smash{{\SetFigFont{6}{7.2}{\familydefault}{\mddefault}{\updefault}{\color[rgb]{0,0,0}$q_{2}$}%
}}}}
\put(1951,-6061){\makebox(0,0)[lb]{\smash{{\SetFigFont{6}{7.2}{\familydefault}{\mddefault}{\updefault}{\color[rgb]{0,0,0}$U_{1}$}%
}}}}
\put(2101,-6961){\makebox(0,0)[lb]{\smash{{\SetFigFont{6}{7.2}{\familydefault}{\mddefault}{\updefault}{\color[rgb]{0,0,0}$O$}%
}}}}
\put(9451,-6211){\makebox(0,0)[lb]{\smash{{\SetFigFont{6}{7.2}{\familydefault}{\mddefault}{\updefault}{\color[rgb]{0,0,0}$q_{1}$}%
}}}}
\put(3901,-1411){\makebox(0,0)[lb]{\smash{{\SetFigFont{6}{7.2}{\familydefault}{\mddefault}{\updefault}{\color[rgb]{0,0,0}$E(z_{3})=E([0,2,3,3])=E(\frac{10}{23})$}%
}}}}
\end{picture}%
      \caption{A digital straight segment of characteristics
        $(10,23,0)$ with an odd slope, taken between origin
        and its second lower leaning point.}
      \label{fig:aop-pattern}
    \end{center}
  \end{figure}

Let us now explain how to compute the \emph{pattern} associated with a
rational slope $z$ in the first octant (i.e. $z=\frac{a}{b}$ with $ 0
\leq a < b$). Horizontal steps are denoted by $0$ and vertical steps
are denoted by $1$. Let us define $E$ a mapping from the set of
positive rational number smaller than one onto the Freeman-code's
words, more precisely $E(z_0) = 0$, $E(z_1) = 0^{u_1}1$ and others
values are expressed recursively:
\begin{eqnarray}
  E(z_{2i+1}) & = & E(z_{2i})^{u_{2i+1}} E(z_{2i-1}), 
  \label{pattern:rec:odd}\\ 
  E(z_{2i}) & = & E(z_{2i - 2}) E(z_{2i-1})^{u_{2i}}.
  \label{pattern:rec:even}
\end{eqnarray}
It has been shown that this mapping constructs the pattern $(a,b)$ for
any rational slope $z=\frac{a}{b}$. \refFigure{fig:aop-pattern}
exemplifies the construction.

There exists other equivalent relations for computing numerators and
denominators (see \cite{Klette04} Chap.~9 and \cite{Voss93} Chap.~4)
and the \emph{splitting formula} can be used to obtain patterns.
However the splitting formula uses two $k$-th convergent with the same
depth, whereas we here use two $k$-th convergent of consecutive
depth. The {\em parity} of a slope is defined as the parity of
the depth of its development in continued fractions. 

\section{Combinatoric view of {\bf DR95} algorithm}
\label{sec:dr95}

Changes in the slope with the {\bf DR95} algorithm occur when weakly
exterior points are added to the segment. We propose here to explain
the different classes of parameters that rule the evolution process,
that is, the characteristics of the straight segment $(a,b,\mu)$, the
numbers of patterns and reversed-pattern $(a,b)$ that constitute it,
the depth of the rational fraction $\frac{a}{b}$, the type of weakly
exterior point that is added (upper or lower) and if it is added to
the right or to the left.

If a digital straight segment with characteristics $(a,b,\mu)$ does
not contain any \emph{pattern} $(a,b)$ then it only contains a
\emph{reversed-pattern} $(a,b)$, that is, two lower leaning points and
one upper leaning point.
\begin{lemma}
  If a digital straight segment does not contain any pattern, then
  there is necessarily one upper leaning point laying on the digital
  path before an upper weakly exterior point. Similarly, if a digital
  straight segment does not contain any reversed-pattern, then there
  is necessarily one lower leaning point laying on the digital path
  before a lower weakly exterior point.
\end{lemma}
\begin{proof} 
  Due to the values of the remainder, one can see that weakly exterior
  points and leaning points are connected
  (\refFigure{fig:pointOrder}). Consider a digital straight line of
  characteristics $(a,b,\mu)$, and let $U$ (resp. $L$) be an upper
  (resp. lower) leaning point of that line. The point $U + (1,-1)$ has
  a remainder of $\mu + a + b$ which means it is a lower weakly
  exterior point. A similar reasoning shows us that $L + (-1,1)$ is
  always an upper weakly exterior point. As a result, leaning points
  and weakly exterior points are ordered in a particular way. We can
  therefore state that before a weakly exterior point lays a leaning
  point of the same type. \qed
\end{proof}

Thus we consider that before adding an upper weakly exterior point to
a digital straight segment, $\delta$ is always greater than or equal
to one. Similarly before adding a lower weakly exterior point to a
digital segment, $\delta'$ is always greater than or equal to one. Let
us state precisely the evolution of the slope of a segment when adding
an upper weakly exterior point to its left or its right.
\begin{proposition} \label{prop:dr95}
  The evolution of the slope of a DSS recognized with the {\bf DR95}
  algorithm depends of the parity of its depth, the type of weakly
  exterior point added and the side where it is added. This process can be
  summed up as follows:
  \begin{itemize}
  \item slope with even depth $[0,u_{1},\ldots,u_{2i}]$, $\delta$ pattern(s) and $\delta'$ reversed-pattern(s): \\[1mm]
    \begin{tabular}{|c|c|c|}
      \cline{2-3}
        \multicolumn{1}{c|}{} &    Left side & Right side \\
      \hline
      Upper weakly exterior &  $[0,u_{1},\ldots,u_{2i}-1,1,\delta]$ & $[0,u_{1},\ldots,u_{2i},\delta]$ \\
      \hline
      Lower weakly  exterior & $[0,u_{1},\ldots,u_{2i},\delta']$ & $[0,u_{1},\ldots,u_{2i}-1,1,\delta']$ \\
      \hline
    \end{tabular}
    \vspace{1mm}

  \item slope with odd depth  $[0,u_{1},\ldots,u_{2i+1}]$, $\delta$ pattern(s) and $\delta'$ reversed-pattern(s): \\[1mm]
    \begin{tabular}{|c|c|c|}
      \cline{2-3}
        \multicolumn{1}{c|}{}  &    Left side & Right side \\
      \hline
       Upper weakly exterior &  $[0,u_{1},\ldots,u_{2i+1},\delta]$ & $[0,u_{1},\ldots,u_{2i+1}-1,1,\delta]$ \\
      \hline
      Lower weakly exterior & $[0,u_{1},\ldots,u_{2i+1}-1,1,\delta']$ & $[0,u_{1},\ldots,u_{2i+1},\delta']$ \\
      \hline
    \end{tabular}
  \end{itemize}
\end{proposition}
\begin{proof}
  We give the proof in the case of an even slope when an upper weakly
  exterior point is added to the right side. Other cases are
  deduced with a similar reasoning or considering the segment
  upside-down.

  Consider we have $\delta \geq 1$. Let $U_{L}$ and $U_{R}$ be the
  leftmost and rightmost upper leaning point of the DSS. We choose
  $U_{L}$ as the origin.  Let $\frac{p_{2i}}{q_{2i}} =
  [0;u_{1},\ldots,u_{2i}]$.

  The added point $U'$ has a remainder equal to $-1$.
  \Equ{pattern:rec:dif} indicates that $q_{2i-1}$ and $ p_{2i-1}$ are
  the smallest positive Bezout coefficient verifying $p_{2i}x -
  q_{2i}y = -1$. As a result: ${ \bf U_{R}U'} = (q_{2i-1},p_{2i-1})$.

  From the recognition algorithm {\bf DR95} the slope of the segment
  $\PTS{U_{L}}{U'}$ is given by the vector $\VEC{U_{L}U'}$. Since
  $\VEC{U_{L}U'} = \VEC{U_{L}U_{R}} + \VEC{U_{R}U'}$, $\VEC{U_{L}U'}$
  equals $\delta (q_{2i},p_{2i}) + (q_{2i-1},p_{2i-1})$. From
  \Equ{pattern:rec:num} and (\ref{pattern:rec:den}) this slope equal
  $\frac{p_{2i+1}}{q_{2i+1}} = \linebreak[4] [0,u_{1},\ldots,u_{2i},\delta]$. If
  $\delta$ equals one then $[0,u_{1},\ldots,u_{2i},1] =
  [0,u_{1},\ldots,u_{2i}+1]$. \qed 
\end{proof}

The slope depth of a DSS when adding a weakly exterior point remains
the same or is increased by one or two.

\section{Connection with the Stern-Brocot tree}
\label{sec:sbt}

We now show that the evolution of the DSS parameters during the
recognition process, analytically given in \refProposition{prop:dr95},
can be traced on the Stern-Brocot tree and has then a more intuitive
interpretation. The Stern-Brocot tree represents all positive rational
fractions. It was already observed that the {\bf DR95} recognition
algorithm has some connection with it \cite{Debled95t}. More
precisely the successive values of the slope taken by a segment are
deeper and deeper nodes of this tree. Note that this tree has other
connections with discrete geometry, like determining the minimal
characteristics of the intersection of two digital straight lines
\cite{Sivignon03,Sivignon04}.

\begin{figure}[t] 
  \begin{center}
    \begin{picture}(0,0)%
\includegraphics{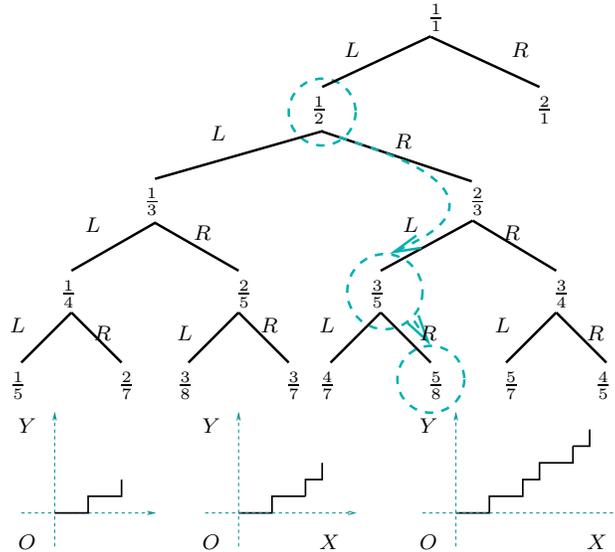}%
\end{picture}%
\setlength{\unitlength}{1381sp}%
\begingroup\makeatletter\ifx\SetFigFont\undefined%
\gdef\SetFigFont#1#2#3#4#5{%
  \reset@font\fontsize{#1}{#2pt}%
  \fontfamily{#3}\fontseries{#4}\fontshape{#5}%
  \selectfont}%
\fi\endgroup%
\begin{picture}(10996,9832)(451,-12155)
\put(6451,-3211){\makebox(0,0)[lb]{\smash{{\SetFigFont{8}{9.6}{\familydefault}{\mddefault}{\updefault}{\color[rgb]{0,0,0}$L$}%
}}}}
\put(601,-12061){\makebox(0,0)[lb]{\smash{{\SetFigFont{8}{9.6}{\familydefault}{\mddefault}{\updefault}{\color[rgb]{0,0,0}$O$}%
}}}}
\put(601,-9961){\makebox(0,0)[lb]{\smash{{\SetFigFont{8}{9.6}{\familydefault}{\mddefault}{\updefault}{\color[rgb]{0,0,0}$Y$}%
}}}}
\put(7801,-12061){\makebox(0,0)[lb]{\smash{{\SetFigFont{8}{9.6}{\familydefault}{\mddefault}{\updefault}{\color[rgb]{0,0,0}$O$}%
}}}}
\put(7801,-9961){\makebox(0,0)[lb]{\smash{{\SetFigFont{8}{9.6}{\familydefault}{\mddefault}{\updefault}{\color[rgb]{0,0,0}$Y$}%
}}}}
\put(10801,-12061){\makebox(0,0)[lb]{\smash{{\SetFigFont{8}{9.6}{\familydefault}{\mddefault}{\updefault}{\color[rgb]{0,0,0}$X$}%
}}}}
\put(6001,-12061){\makebox(0,0)[lb]{\smash{{\SetFigFont{8}{9.6}{\familydefault}{\mddefault}{\updefault}{\color[rgb]{0,0,0}$X$}%
}}}}
\put(3901,-12061){\makebox(0,0)[lb]{\smash{{\SetFigFont{8}{9.6}{\familydefault}{\mddefault}{\updefault}{\color[rgb]{0,0,0}$O$}%
}}}}
\put(3901,-9961){\makebox(0,0)[lb]{\smash{{\SetFigFont{8}{9.6}{\familydefault}{\mddefault}{\updefault}{\color[rgb]{0,0,0}$Y$}%
}}}}
\put(2851,-5911){\makebox(0,0)[lb]{\smash{{\SetFigFont{8}{9.6}{\familydefault}{\mddefault}{\updefault}{\color[rgb]{0,0,0}$\frac{1}{3}$}%
}}}}
\put(1351,-7561){\makebox(0,0)[lb]{\smash{{\SetFigFont{8}{9.6}{\familydefault}{\mddefault}{\updefault}{\color[rgb]{0,0,0}$\frac{1}{4}$}%
}}}}
\put(451,-9211){\makebox(0,0)[lb]{\smash{{\SetFigFont{8}{9.6}{\familydefault}{\mddefault}{\updefault}{\color[rgb]{0,0,0}$\frac{1}{5}$}%
}}}}
\put(3451,-9211){\makebox(0,0)[lb]{\smash{{\SetFigFont{8}{9.6}{\familydefault}{\mddefault}{\updefault}{\color[rgb]{0,0,0}$\frac{3}{8}$}%
}}}}
\put(6001,-9211){\makebox(0,0)[lb]{\smash{{\SetFigFont{8}{9.6}{\familydefault}{\mddefault}{\updefault}{\color[rgb]{0,0,0}$\frac{4}{7}$}%
}}}}
\put(5851,-4261){\makebox(0,0)[lb]{\smash{{\SetFigFont{8}{9.6}{\familydefault}{\mddefault}{\updefault}{\color[rgb]{0,0,0}$\frac{1}{2}$}%
}}}}
\put(6901,-7561){\makebox(0,0)[lb]{\smash{{\SetFigFont{8}{9.6}{\familydefault}{\mddefault}{\updefault}{\color[rgb]{0,0,0}$\frac{3}{5}$}%
}}}}
\put(8701,-5911){\makebox(0,0)[lb]{\smash{{\SetFigFont{8}{9.6}{\familydefault}{\mddefault}{\updefault}{\color[rgb]{0,0,0}$\frac{2}{3}$}%
}}}}
\put(4501,-7561){\makebox(0,0)[lb]{\smash{{\SetFigFont{8}{9.6}{\familydefault}{\mddefault}{\updefault}{\color[rgb]{0,0,0}$\frac{2}{5}$}%
}}}}
\put(10201,-7561){\makebox(0,0)[lb]{\smash{{\SetFigFont{8}{9.6}{\familydefault}{\mddefault}{\updefault}{\color[rgb]{0,0,0}$\frac{3}{4}$}%
}}}}
\put(2401,-9211){\makebox(0,0)[lb]{\smash{{\SetFigFont{8}{9.6}{\familydefault}{\mddefault}{\updefault}{\color[rgb]{0,0,0}$\frac{2}{7}$}%
}}}}
\put(5401,-9211){\makebox(0,0)[lb]{\smash{{\SetFigFont{8}{9.6}{\familydefault}{\mddefault}{\updefault}{\color[rgb]{0,0,0}$\frac{3}{7}$}%
}}}}
\put(7951,-9211){\makebox(0,0)[lb]{\smash{{\SetFigFont{8}{9.6}{\familydefault}{\mddefault}{\updefault}{\color[rgb]{0,0,0}$\frac{5}{8}$}%
}}}}
\put(9301,-9211){\makebox(0,0)[lb]{\smash{{\SetFigFont{8}{9.6}{\familydefault}{\mddefault}{\updefault}{\color[rgb]{0,0,0}$\frac{5}{7}$}%
}}}}
\put(10951,-9211){\makebox(0,0)[lb]{\smash{{\SetFigFont{8}{9.6}{\familydefault}{\mddefault}{\updefault}{\color[rgb]{0,0,0}$\frac{4}{5}$}%
}}}}
\put(7951,-2611){\makebox(0,0)[lb]{\smash{{\SetFigFont{8}{9.6}{\familydefault}{\mddefault}{\updefault}{\color[rgb]{0,0,0}$\frac{1}{1}$}%
}}}}
\put(9901,-4261){\makebox(0,0)[lb]{\smash{{\SetFigFont{8}{9.6}{\familydefault}{\mddefault}{\updefault}{\color[rgb]{0,0,0}$\frac{2}{1}$}%
}}}}
\put(9451,-3211){\makebox(0,0)[lb]{\smash{{\SetFigFont{8}{9.6}{\familydefault}{\mddefault}{\updefault}{\color[rgb]{0,0,0}$R$}%
}}}}
\put(4051,-4711){\makebox(0,0)[lb]{\smash{{\SetFigFont{8}{9.6}{\familydefault}{\mddefault}{\updefault}{\color[rgb]{0,0,0}$L$}%
}}}}
\put(1801,-6361){\makebox(0,0)[lb]{\smash{{\SetFigFont{8}{9.6}{\familydefault}{\mddefault}{\updefault}{\color[rgb]{0,0,0}$L$}%
}}}}
\put(451,-8161){\makebox(0,0)[lb]{\smash{{\SetFigFont{8}{9.6}{\familydefault}{\mddefault}{\updefault}{\color[rgb]{0,0,0}$L$}%
}}}}
\put(7351,-4861){\makebox(0,0)[lb]{\smash{{\SetFigFont{8}{9.6}{\familydefault}{\mddefault}{\updefault}{\color[rgb]{0,0,0}$R$}%
}}}}
\put(3751,-6511){\makebox(0,0)[lb]{\smash{{\SetFigFont{8}{9.6}{\familydefault}{\mddefault}{\updefault}{\color[rgb]{0,0,0}$R$}%
}}}}
\put(4951,-8161){\makebox(0,0)[lb]{\smash{{\SetFigFont{8}{9.6}{\familydefault}{\mddefault}{\updefault}{\color[rgb]{0,0,0}$R$}%
}}}}
\put(7801,-8311){\makebox(0,0)[lb]{\smash{{\SetFigFont{8}{9.6}{\familydefault}{\mddefault}{\updefault}{\color[rgb]{0,0,0}$R$}%
}}}}
\put(9301,-6511){\makebox(0,0)[lb]{\smash{{\SetFigFont{8}{9.6}{\familydefault}{\mddefault}{\updefault}{\color[rgb]{0,0,0}$R$}%
}}}}
\put(10801,-8311){\makebox(0,0)[lb]{\smash{{\SetFigFont{8}{9.6}{\familydefault}{\mddefault}{\updefault}{\color[rgb]{0,0,0}$R$}%
}}}}
\put(9151,-8161){\makebox(0,0)[lb]{\smash{{\SetFigFont{8}{9.6}{\familydefault}{\mddefault}{\updefault}{\color[rgb]{0,0,0}$L$}%
}}}}
\put(6001,-8161){\makebox(0,0)[lb]{\smash{{\SetFigFont{8}{9.6}{\familydefault}{\mddefault}{\updefault}{\color[rgb]{0,0,0}$L$}%
}}}}
\put(3451,-8311){\makebox(0,0)[lb]{\smash{{\SetFigFont{8}{9.6}{\familydefault}{\mddefault}{\updefault}{\color[rgb]{0,0,0}$L$}%
}}}}
\put(7501,-6361){\makebox(0,0)[lb]{\smash{{\SetFigFont{8}{9.6}{\familydefault}{\mddefault}{\updefault}{\color[rgb]{0,0,0}$L$}%
}}}}
\put(1951,-8311){\makebox(0,0)[lb]{\smash{{\SetFigFont{8}{9.6}{\familydefault}{\mddefault}{\updefault}{\color[rgb]{0,0,0}$R$}%
}}}}
\end{picture}%
    \caption{Evolution of the slope of a digital straight
      segment. Successive values are $\frac{1}{2}$, $\frac{3}{5}$
      and $\frac{5}{8}$. Modifications are triggered by the addition
      of an upper weakly exterior point. The successive slope depths
      are one, three and four. 
      Nodes with depth one are $\frac{1}{1}$,$\frac{1}{2}$,$\frac{1}{3}$,$\frac{1}{4}$,$\frac{1}{5}$, 
      nodes with depth two are $\frac{2}{3}$,$\frac{3}{4}$,$\frac{4}{5}$,$\frac{2}{5}$,$\frac{3}{7}$,$\frac{2}{7}$
      nodes with depth three are $\frac{3}{5}$, $\frac{4}{7}$,$\frac{3}{8}$,$\frac{5}{7}$ and 
      node with depth four is $\frac{5}{8}$.
      \label{fig:EvolStern-Brocot} }
  \end{center}
\end{figure}

The Stern-Brocot tree is a binary tree constructed by starting with
the fractions $\frac{0}{1}$ and $\frac{1}{0}$ and iteratively
inserting $\frac{m+m'}{n+n'}$ between each two adjacent fractions
$\frac{m}{n}$ and $\frac{m'}{n'}$ (\refFigure{fig:EvolStern-Brocot}).
Any node with a value between $0$ (excluded) and $1$ (included) is
obtained by finite successive moves from the $\frac{1}{1}$ node. Those
moves can be of two types: $L$ is a move toward the left child, $R$ is
a move toward the right child. Those moves determine the type of node
when they end paths: even (resp. odd) nodes end with a $R$ (resp. $L$)
move. It is known that those nodes have a development in continued
fraction, and is such that:
\begin{itemize}
\item even nodes:\\
$[0, u_{1}, \ldots, u_{2k}, u_{2k+1}, \ldots, u_{2i-1},u_{2i}] \equiv R^{0}L^{u_{1}} \ldots R^{u_{2k}}L^{u_{2k+1}} \ldots L^{u_{2i-1}}R^{u_{2i}-1}$ 
\item odd nodes: \\
$[0, u_{1}, \ldots ,u_{2k}, u_{2k+1}, \ldots, u_{2i},u_{2i+1}] \equiv R^{0}L^{u_{1}} \ldots R^{u_{2k}}L^{u_{2k+1}} \ldots R^{u_{2i}}L^{u_{2i+1}-1}$
\end{itemize} 
Of course odd nodes have an odd depth in their development in
continued fractions, similarly even nodes have an even depth. When
descending the tree the depth of a child changes if the move used to
reach it differs from the last move used to reach its father.
Consider the node $\frac{1}{2}$ whose depth equals one, the last move
used to reach it is a $L$ move, its left child $\frac{1}{3}$ has the
same depth.  The right child of $\frac{1}{2}$ is obtained by the
successive moves $R^{0}L^{1}R^{1}$ and has a depth that equals two.
We can classify the nodes of the Stern-Brocot tree according to the
depth of their development in continued fraction
(\refFigure{fig:EvolStern-Brocot}).

Nodes of this tree may also be viewed as slopes of a digital straight
segment being recognized. As we consider rational slopes between zero
and one, we only consider nodes whose value is between zero and one.
All those nodes are derived (except for the zero node) from the
$\frac{1}{1}$ node with a $L$ move first. This implies that $u_{0}$
equals zero. It is possible to trace the slope evolution of a digital
straight segment during recognition on the Stern-Brocot tree, as
exemplified in \refFigure{fig:EvolStern-Brocot}.

\begin{figure}[htp]
  \begin{center}
    \hspace{-2.0cm} 
\begin{picture}(0,0)%
\includegraphics{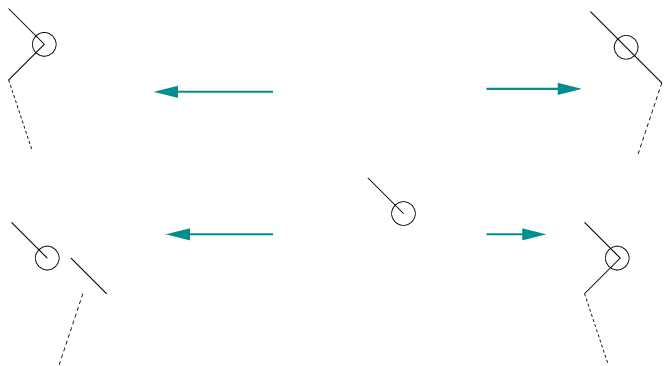}%
\end{picture}%
\setlength{\unitlength}{750sp}%
\begingroup\makeatletter\ifx\SetFigFont\undefined%
\gdef\SetFigFont#1#2#3#4#5{%
  \reset@font\fontsize{#1}{#2pt}%
  \fontfamily{#3}\fontseries{#4}\fontshape{#5}%
  \selectfont}%
\fi\endgroup%
\begin{picture}(17715,10435)(-5174,-12683)
\put(-3599,-9361){\makebox(0,0)[lb]{\smash{{\SetFigFont{5}{6.0}{\familydefault}{\mddefault}{\updefault}{\color[rgb]{0,0,0}$R$}%
}}}}
\put(11026,-4036){\makebox(0,0)[lb]{\smash{{\SetFigFont{5}{6.0}{\familydefault}{\mddefault}{\updefault}{\color[rgb]{0,0,0}$R$}%
}}}}
\put(5401,-8236){\makebox(0,0)[lb]{\smash{{\SetFigFont{5}{6.0}{\familydefault}{\mddefault}{\updefault}{\color[rgb]{0,0,0}$R$}%
}}}}
\put(2701,-9361){\makebox(0,0)[lb]{\smash{{\SetFigFont{5}{6.0}{\familydefault}{\mddefault}{\updefault}{\color[rgb]{0,0,0}$L'$}%
}}}}
\put(6901,-9361){\makebox(0,0)[lb]{\smash{{\SetFigFont{5}{6.0}{\familydefault}{\mddefault}{\updefault}{\color[rgb]{0,0,0}$L'$}%
}}}}
\put(2626,-5761){\makebox(0,0)[lb]{\smash{{\SetFigFont{5}{6.0}{\familydefault}{\mddefault}{\updefault}{\color[rgb]{0,0,0}$U'$}%
}}}}
\put(6901,-5761){\makebox(0,0)[lb]{\smash{{\SetFigFont{5}{6.0}{\familydefault}{\mddefault}{\updefault}{\color[rgb]{0,0,0}$U'$}%
}}}}
\put(-3674,-3961){\makebox(0,0)[lb]{\smash{{\SetFigFont{5}{6.0}{\familydefault}{\mddefault}{\updefault}{\color[rgb]{0,0,0}$R$}%
}}}}
\put(-3974,-6061){\makebox(0,0)[lb]{\smash{{\SetFigFont{5}{6.0}{\familydefault}{\mddefault}{\updefault}{\color[rgb]{0,0,0}$R^{\delta -1}$}%
}}}}
\put(-4649,-4786){\makebox(0,0)[lb]{\smash{{\SetFigFont{5}{6.0}{\familydefault}{\mddefault}{\updefault}{\color[rgb]{0,0,0}$L$}%
}}}}
\put(-2699,-10261){\makebox(0,0)[lb]{\smash{{\SetFigFont{5}{6.0}{\familydefault}{\mddefault}{\updefault}{\color[rgb]{0,0,0}$R$}%
}}}}
\put(-2399,-12061){\makebox(0,0)[lb]{\smash{{\SetFigFont{5}{6.0}{\familydefault}{\mddefault}{\updefault}{\color[rgb]{0,0,0}$L^{\delta' -1}$}%
}}}}
\put(11926,-4936){\makebox(0,0)[lb]{\smash{{\SetFigFont{5}{6.0}{\familydefault}{\mddefault}{\updefault}{\color[rgb]{0,0,0}$R$}%
}}}}
\put(12226,-6736){\makebox(0,0)[lb]{\smash{{\SetFigFont{5}{6.0}{\familydefault}{\mddefault}{\updefault}{\color[rgb]{0,0,0}$L^{\delta -1}$}%
}}}}
\put(10876,-9361){\makebox(0,0)[lb]{\smash{{\SetFigFont{5}{6.0}{\familydefault}{\mddefault}{\updefault}{\color[rgb]{0,0,0}$R$}%
}}}}
\put(9976,-10261){\makebox(0,0)[lb]{\smash{{\SetFigFont{5}{6.0}{\familydefault}{\mddefault}{\updefault}{\color[rgb]{0,0,0}$L$}%
}}}}
\put(10576,-11461){\makebox(0,0)[lb]{\smash{{\SetFigFont{5}{6.0}{\familydefault}{\mddefault}{\updefault}{\color[rgb]{0,0,0}$R^{\delta' -1}$}%
}}}}
\put(-5174,-2536){\makebox(0,0)[lb]{\smash{{\SetFigFont{5}{6.0}{\familydefault}{\mddefault}{\updefault}{\color[rgb]{0,0,0}$[0,u_{1}, \ldots, u_{2i}-1, 1, \delta]$}%
}}}}
\put(-5099,-7936){\makebox(0,0)[lb]{\smash{{\SetFigFont{5}{6.0}{\familydefault}{\mddefault}{\updefault}{\color[rgb]{0,0,0}$[0,u_{1},\ldots,u_{2i},\delta']$}%
}}}}
\put(2401,-6436){\makebox(0,0)[lb]{\smash{{\SetFigFont{5}{6.0}{\familydefault}{\mddefault}{\updefault}{\color[rgb]{0,0,0}$[0,u_{1},\ldots,u_{2i}]$}%
}}}}
\put(9376,-7936){\makebox(0,0)[lb]{\smash{{\SetFigFont{5}{6.0}{\familydefault}{\mddefault}{\updefault}{\color[rgb]{0,0,0}$[0,u_{1}, \ldots, u_{2i}-1, 1, \delta']$}%
}}}}
\put(9526,-2611){\makebox(0,0)[lb]{\smash{{\SetFigFont{5}{6.0}{\familydefault}{\mddefault}{\updefault}{\color[rgb]{0,0,0}$[0,u_{1},\ldots,u_{2i},\delta]$}%
}}}}
\put(2401,-7186){\makebox(0,0)[lb]{\smash{{\SetFigFont{5}{6.0}{\familydefault}{\mddefault}{\updefault}{\color[rgb]{0,0,0}$R^{0}L^{u_{1}} \ldots R^{u_{2i} -1}$}%
}}}}
\put(-5174,-3286){\makebox(0,0)[lb]{\smash{{\SetFigFont{5}{6.0}{\familydefault}{\mddefault}{\updefault}{\color[rgb]{0,0,0}$R^{0}L^{u_{1}} \ldots R^{u_{2i} -1}L^{1}R^{\delta -1}$}%
}}}}
\put(9526,-3286){\makebox(0,0)[lb]{\smash{{\SetFigFont{5}{6.0}{\familydefault}{\mddefault}{\updefault}{\color[rgb]{0,0,0}$R^{0}L^{u_{1}} \ldots R^{u_{2i} }L^{\delta -1}$}%
}}}}
\put(9376,-8611){\makebox(0,0)[lb]{\smash{{\SetFigFont{5}{6.0}{\familydefault}{\mddefault}{\updefault}{\color[rgb]{0,0,0}$R^{0}L^{u_{1}} \ldots R^{u_{2i} -1}L^{1}R^{\delta' -1}$}%
}}}}
\put(-5099,-8611){\makebox(0,0)[lb]{\smash{{\SetFigFont{5}{6.0}{\familydefault}{\mddefault}{\updefault}{\color[rgb]{0,0,0}$R^{0}L^{u_{1}} \ldots R^{u_{2i} }L^{\delta' -1}$}%
}}}}
\end{picture}%
\end{center}
\vspace{-0.2cm}
\begin{center}
  \hspace{-2.0cm} 
\begin{picture}(0,0)%
\includegraphics{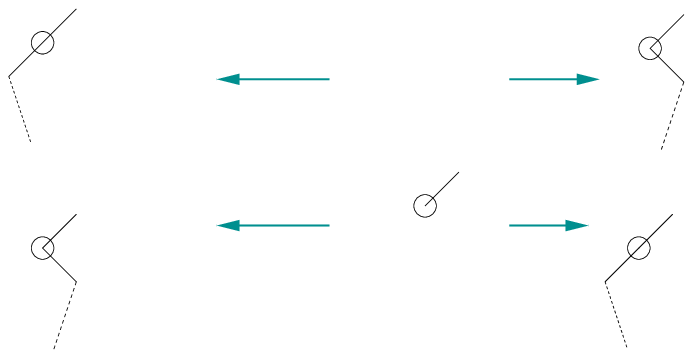}%
\end{picture}%
\setlength{\unitlength}{710sp}%
\begingroup\makeatletter\ifx\SetFigFont\undefined%
\gdef\SetFigFont#1#2#3#4#5{%
  \reset@font\fontsize{#1}{#2pt}%
  \fontfamily{#3}\fontseries{#4}\fontshape{#5}%
  \selectfont}%
\fi\endgroup%
\begin{picture}(19215,10660)(-7349,-12683)
\put(3901,-8236){\makebox(0,0)[lb]{\smash{{\SetFigFont{5}{6.0}{\familydefault}{\mddefault}{\updefault}{\color[rgb]{0,0,0}$L$}%
}}}}
\put(2251,-5461){\makebox(0,0)[lb]{\smash{{\SetFigFont{5}{6.0}{\familydefault}{\mddefault}{\updefault}{\color[rgb]{0,0,0}$U'$}%
}}}}
\put(2401,-9361){\makebox(0,0)[lb]{\smash{{\SetFigFont{5}{6.0}{\familydefault}{\mddefault}{\updefault}{\color[rgb]{0,0,0}$L'$}%
}}}}
\put(6001,-9361){\makebox(0,0)[lb]{\smash{{\SetFigFont{5}{6.0}{\familydefault}{\mddefault}{\updefault}{\color[rgb]{0,0,0}$L'$}%
}}}}
\put(6001,-5461){\makebox(0,0)[lb]{\smash{{\SetFigFont{5}{6.0}{\familydefault}{\mddefault}{\updefault}{\color[rgb]{0,0,0}$U'$}%
}}}}
\put(1951,-6136){\makebox(0,0)[lb]{\smash{{\SetFigFont{5}{6.0}{\familydefault}{\mddefault}{\updefault}{\color[rgb]{0,0,0}$[0,u_{1},\ldots,u_{2i+1}]$}%
}}}}
\put(1951,-6961){\makebox(0,0)[lb]{\smash{{\SetFigFont{5}{6.0}{\familydefault}{\mddefault}{\updefault}{\color[rgb]{0,0,0}$R^{0}L^{u_{1}} \ldots L^{u_{2i+1} -1}$}%
}}}}
\put(11551,-6736){\makebox(0,0)[lb]{\smash{{\SetFigFont{5}{6.0}{\familydefault}{\mddefault}{\updefault}{\color[rgb]{0,0,0}$L^{\delta -1}$}%
}}}}
\put(10351,-4036){\makebox(0,0)[lb]{\smash{{\SetFigFont{5}{6.0}{\familydefault}{\mddefault}{\updefault}{\color[rgb]{0,0,0}$L$}%
}}}}
\put(11251,-4936){\makebox(0,0)[lb]{\smash{{\SetFigFont{5}{6.0}{\familydefault}{\mddefault}{\updefault}{\color[rgb]{0,0,0}$R$}%
}}}}
\put(8851,-2536){\makebox(0,0)[lb]{\smash{{\SetFigFont{5}{6.0}{\familydefault}{\mddefault}{\updefault}{\color[rgb]{0,0,0}$[0,u_{1},\ldots,u_{2i+1}-1,1,\delta]$}%
}}}}
\put(9751,-11461){\makebox(0,0)[lb]{\smash{{\SetFigFont{5}{6.0}{\familydefault}{\mddefault}{\updefault}{\color[rgb]{0,0,0}$R^{\delta' -1}$}%
}}}}
\put(10051,-9361){\makebox(0,0)[lb]{\smash{{\SetFigFont{5}{6.0}{\familydefault}{\mddefault}{\updefault}{\color[rgb]{0,0,0}$L$}%
}}}}
\put(9151,-10261){\makebox(0,0)[lb]{\smash{{\SetFigFont{5}{6.0}{\familydefault}{\mddefault}{\updefault}{\color[rgb]{0,0,0}$L$}%
}}}}
\put(8851,-7861){\makebox(0,0)[lb]{\smash{{\SetFigFont{5}{6.0}{\familydefault}{\mddefault}{\updefault}{\color[rgb]{0,0,0}$[0,u_{1},\ldots,u_{2i+1},\delta']$}%
}}}}
\put(-6149,-5986){\makebox(0,0)[lb]{\smash{{\SetFigFont{5}{6.0}{\familydefault}{\mddefault}{\updefault}{\color[rgb]{0,0,0}$R^{\delta -1}$}%
}}}}
\put(-5849,-3886){\makebox(0,0)[lb]{\smash{{\SetFigFont{5}{6.0}{\familydefault}{\mddefault}{\updefault}{\color[rgb]{0,0,0}$L$}%
}}}}
\put(-6749,-4786){\makebox(0,0)[lb]{\smash{{\SetFigFont{5}{6.0}{\familydefault}{\mddefault}{\updefault}{\color[rgb]{0,0,0}$L$}%
}}}}
\put(-7349,-2311){\makebox(0,0)[lb]{\smash{{\SetFigFont{5}{6.0}{\familydefault}{\mddefault}{\updefault}{\color[rgb]{0,0,0}$[0,u_{1},\ldots,u_{2i+1},\delta]$}%
}}}}
\put(-4649,-12061){\makebox(0,0)[lb]{\smash{{\SetFigFont{5}{6.0}{\familydefault}{\mddefault}{\updefault}{\color[rgb]{0,0,0}$L^{\delta' -1}$}%
}}}}
\put(-5849,-9361){\makebox(0,0)[lb]{\smash{{\SetFigFont{5}{6.0}{\familydefault}{\mddefault}{\updefault}{\color[rgb]{0,0,0}$L$}%
}}}}
\put(-4949,-10261){\makebox(0,0)[lb]{\smash{{\SetFigFont{5}{6.0}{\familydefault}{\mddefault}{\updefault}{\color[rgb]{0,0,0}$R$}%
}}}}
\put(-7349,-7861){\makebox(0,0)[lb]{\smash{{\SetFigFont{5}{6.0}{\familydefault}{\mddefault}{\updefault}{\color[rgb]{0,0,0}$[0,u_{1},\ldots,u_{2i+1}-1,1,\delta']$}%
}}}}
\put(8851,-3361){\makebox(0,0)[lb]{\smash{{\SetFigFont{5}{6.0}{\familydefault}{\mddefault}{\updefault}{\color[rgb]{0,0,0}$R^{0}L^{u_{1}} \ldots L^{u_{2i+1}-1 }R^{1}L^{\delta -1}$}%
}}}}
\put(8851,-8611){\makebox(0,0)[lb]{\smash{{\SetFigFont{5}{6.0}{\familydefault}{\mddefault}{\updefault}{\color[rgb]{0,0,0}$R^{0}L^{u_{1}} \ldots L^{u_{2i+1}}R^{\delta' -1}$}%
}}}}
\put(-7349,-8686){\makebox(0,0)[lb]{\smash{{\SetFigFont{5}{6.0}{\familydefault}{\mddefault}{\updefault}{\color[rgb]{0,0,0}$R^{0}L^{u_{1}} \ldots L^{u_{2i+1}-1 }R^{1}L^{\delta' -1}$}%
}}}}
\put(-7349,-3061){\makebox(0,0)[lb]{\smash{{\SetFigFont{5}{6.0}{\familydefault}{\mddefault}{\updefault}{\color[rgb]{0,0,0}$R^{0}L^{u_{1}} \ldots L^{u_{2i+1}}R^{\delta -1}$}%
}}}}
\end{picture}%
  \end{center}
  \caption{Evolution from a digital straight segment with even slope
    (top) and odd slope (bottom) using the {\bf DR95} algorithm,
    represented in terms of move on the Stern-Brocot tree. The point
    $U'$ (resp. $L'$) is an upper (resp. lower) weakly exterior point
    added to the left (left column) or to the right (right column) of
    the DSS.
    \label{fig:evolOddEvenSlopesStern}}
\end{figure}

Since each node has a particular development in continued fraction,
the results shown in \refProposition{prop:dr95} can be reinterpreted
in terms of descending moves on the Stern-Brocot tree. Thus the slope
evolution of a DSS during recognition can be fully described with $L$
and $R$ moves as shown on \refFigure{fig:evolOddEvenSlopesStern}. We
then see that the number of successive moves of the same type directly
depends on the number of patterns or reversed-patterns.

\begin{figure}[bp]
  \begin{center}
    \begin{picture}(0,0)%
\includegraphics{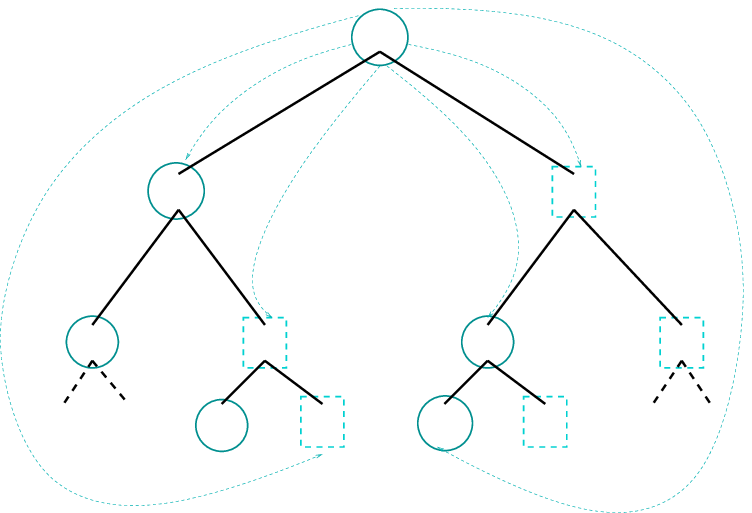}%
\end{picture}%
\setlength{\unitlength}{908sp}%
\begingroup\makeatletter\ifx\SetFigFont\undefined%
\gdef\SetFigFont#1#2#3#4#5{%
  \reset@font\fontsize{#1}{#2pt}%
  \fontfamily{#3}\fontseries{#4}\fontshape{#5}%
  \selectfont}%
\fi\endgroup%
\begin{picture}(15525,10549)(-1929,-11041)
\put(1651,-4411){\makebox(0,0)[lb]{\smash{{\SetFigFont{6}{7.2}{\familydefault}{\mddefault}{\updefault}{\color[rgb]{0,0,0}$\frac{1}{3}$}%
}}}}
\put(3451,-7561){\makebox(0,0)[lb]{\smash{{\SetFigFont{6}{7.2}{\familydefault}{\mddefault}{\updefault}{\color[rgb]{0,0,0}$\frac{2}{5}$}%
}}}}
\put(12151,-7561){\makebox(0,0)[lb]{\smash{{\SetFigFont{6}{7.2}{\familydefault}{\mddefault}{\updefault}{\color[rgb]{0,0,0}$\frac{3}{4}$}%
}}}}
\put(5851,-1111){\makebox(0,0)[lb]{\smash{{\SetFigFont{6}{7.2}{\familydefault}{\mddefault}{\updefault}{\color[rgb]{0,0,0}$\frac{1}{2}$}%
}}}}
\put(7201,-9211){\makebox(0,0)[lb]{\smash{{\SetFigFont{6}{7.2}{\familydefault}{\mddefault}{\updefault}{\color[rgb]{0,0,0}$\frac{4}{7}$}%
}}}}
\put(2551,-9211){\makebox(0,0)[lb]{\smash{{\SetFigFont{6}{7.2}{\familydefault}{\mddefault}{\updefault}{\color[rgb]{0,0,0}$\frac{3}{8}$}%
}}}}
\put(4651,-9211){\makebox(0,0)[lb]{\smash{{\SetFigFont{6}{7.2}{\familydefault}{\mddefault}{\updefault}{\color[rgb]{0,0,0}$\frac{3}{7}$}%
}}}}
\put(9301,-9211){\makebox(0,0)[lb]{\smash{{\SetFigFont{6}{7.2}{\familydefault}{\mddefault}{\updefault}{\color[rgb]{0,0,0}$\frac{5}{8}$}%
}}}}
\put(9901,-4411){\makebox(0,0)[lb]{\smash{{\SetFigFont{6}{7.2}{\familydefault}{\mddefault}{\updefault}{\color[rgb]{0,0,0}$\frac{2}{3}$}%
}}}}
\put(2851,-1561){\makebox(0,0)[lb]{\smash{{\SetFigFont{6}{7.2}{\familydefault}{\mddefault}{\updefault}{\color[rgb]{0,0,0}$\delta'=$}%
}}}}
\put(2851,-2161){\makebox(0,0)[lb]{\smash{{\SetFigFont{6}{7.2}{\familydefault}{\mddefault}{\updefault}{\color[rgb]{0,0,0}$\delta =$}%
}}}}
\put(9451,-1561){\makebox(0,0)[lb]{\smash{{\SetFigFont{6}{7.2}{\familydefault}{\mddefault}{\updefault}{\color[rgb]{0,0,0}$\delta'=$}%
}}}}
\put(9451,-2161){\makebox(0,0)[lb]{\smash{{\SetFigFont{6}{7.2}{\familydefault}{\mddefault}{\updefault}{\color[rgb]{0,0,0}$\delta =$}%
}}}}
\put(4501,-5161){\makebox(0,0)[lb]{\smash{{\SetFigFont{6}{7.2}{\familydefault}{\mddefault}{\updefault}{\color[rgb]{0,0,0}$\delta'=$}%
}}}}
\put(4501,-5761){\makebox(0,0)[lb]{\smash{{\SetFigFont{6}{7.2}{\familydefault}{\mddefault}{\updefault}{\color[rgb]{0,0,0}$\delta =$}%
}}}}
\put(1951,-10261){\makebox(0,0)[lb]{\smash{{\SetFigFont{6}{7.2}{\familydefault}{\mddefault}{\updefault}{\color[rgb]{0,0,0}$\delta'=$}%
}}}}
\put(1951,-10861){\makebox(0,0)[lb]{\smash{{\SetFigFont{6}{7.2}{\familydefault}{\mddefault}{\updefault}{\color[rgb]{0,0,0}$\delta =$}%
}}}}
\put(7951,-10261){\makebox(0,0)[lb]{\smash{{\SetFigFont{6}{7.2}{\familydefault}{\mddefault}{\updefault}{\color[rgb]{0,0,0}$\delta'=$}%
}}}}
\put(7951,-10861){\makebox(0,0)[lb]{\smash{{\SetFigFont{6}{7.2}{\familydefault}{\mddefault}{\updefault}{\color[rgb]{0,0,0}$\delta =$}%
}}}}
\put(8326,-7561){\makebox(0,0)[lb]{\smash{{\SetFigFont{6}{7.2}{\familydefault}{\mddefault}{\updefault}{\color[rgb]{0,0,0}$\frac{3}{5}$}%
}}}}
\put(-149,-7561){\makebox(0,0)[lb]{\smash{{\SetFigFont{6}{7.2}{\familydefault}{\mddefault}{\updefault}{\color[rgb]{0,0,0}$\frac{1}{4}$}%
}}}}
\put(7501,-4561){\makebox(0,0)[lb]{\smash{{\SetFigFont{6}{7.2}{\familydefault}{\mddefault}{\updefault}{\color[rgb]{0,0,0}$\delta =$}%
}}}}
\put(7501,-3961){\makebox(0,0)[lb]{\smash{{\SetFigFont{6}{7.2}{\familydefault}{\mddefault}{\updefault}{\color[rgb]{0,0,0}$\delta'=$}%
}}}}
\put(3901,-1561){\makebox(0,0)[lb]{\smash{{\SetFigFont{6}{7.2}{\familydefault}{\mddefault}{\updefault}{\color[rgb]{0,0,0}$1$}%
}}}}
\put(3901,-2161){\makebox(0,0)[lb]{\smash{{\SetFigFont{6}{7.2}{\familydefault}{\mddefault}{\updefault}{\color[rgb]{0,0,0}$1$}%
}}}}
\put(10501,-1561){\makebox(0,0)[lb]{\smash{{\SetFigFont{6}{7.2}{\familydefault}{\mddefault}{\updefault}{\color[rgb]{0,0,0}$1$}%
}}}}
\put(10501,-2161){\makebox(0,0)[lb]{\smash{{\SetFigFont{6}{7.2}{\familydefault}{\mddefault}{\updefault}{\color[rgb]{0,0,0}$1$}%
}}}}
\put(8551,-4561){\makebox(0,0)[lb]{\smash{{\SetFigFont{6}{7.2}{\familydefault}{\mddefault}{\updefault}{\color[rgb]{0,0,0}$2$}%
}}}}
\put(8551,-3961){\makebox(0,0)[lb]{\smash{{\SetFigFont{6}{7.2}{\familydefault}{\mddefault}{\updefault}{\color[rgb]{0,0,0}$2$}%
}}}}
\put(5551,-5161){\makebox(0,0)[lb]{\smash{{\SetFigFont{6}{7.2}{\familydefault}{\mddefault}{\updefault}{\color[rgb]{0,0,0}$2$}%
}}}}
\put(5551,-5761){\makebox(0,0)[lb]{\smash{{\SetFigFont{6}{7.2}{\familydefault}{\mddefault}{\updefault}{\color[rgb]{0,0,0}$2$}%
}}}}
\put(3001,-10261){\makebox(0,0)[lb]{\smash{{\SetFigFont{6}{7.2}{\familydefault}{\mddefault}{\updefault}{\color[rgb]{0,0,0}$3$}%
}}}}
\put(3001,-10861){\makebox(0,0)[lb]{\smash{{\SetFigFont{6}{7.2}{\familydefault}{\mddefault}{\updefault}{\color[rgb]{0,0,0}$3$}%
}}}}
\put(9001,-10261){\makebox(0,0)[lb]{\smash{{\SetFigFont{6}{7.2}{\familydefault}{\mddefault}{\updefault}{\color[rgb]{0,0,0}$3$}%
}}}}
\put(9001,-10861){\makebox(0,0)[lb]{\smash{{\SetFigFont{6}{7.2}{\familydefault}{\mddefault}{\updefault}{\color[rgb]{0,0,0}$3$}%
}}}}
\put(2401,-1561){\makebox(0,0)[lb]{\smash{{\SetFigFont{6}{7.2}{\familydefault}{\mddefault}{\updefault}{\color[rgb]{0,0,0}$\wedge$}%
}}}}
\put(2401,-2161){\makebox(0,0)[lb]{\smash{{\SetFigFont{6}{7.2}{\familydefault}{\mddefault}{\updefault}{\color[rgb]{0,0,0}$\wedge$}%
}}}}
\put(9001,-1561){\makebox(0,0)[lb]{\smash{{\SetFigFont{6}{7.2}{\familydefault}{\mddefault}{\updefault}{\color[rgb]{0,0,0}$\wedge$}%
}}}}
\put(9001,-2161){\makebox(0,0)[lb]{\smash{{\SetFigFont{6}{7.2}{\familydefault}{\mddefault}{\updefault}{\color[rgb]{0,0,0}$\wedge$}%
}}}}
\put(7051,-3961){\makebox(0,0)[lb]{\smash{{\SetFigFont{6}{7.2}{\familydefault}{\mddefault}{\updefault}{\color[rgb]{0,0,0}$\wedge$}%
}}}}
\put(1651,-10261){\makebox(0,0)[lb]{\smash{{\SetFigFont{6}{7.2}{\familydefault}{\mddefault}{\updefault}{\color[rgb]{0,0,0}$\wedge$}%
}}}}
\put(1651,-10861){\makebox(0,0)[lb]{\smash{{\SetFigFont{6}{7.2}{\familydefault}{\mddefault}{\updefault}{\color[rgb]{0,0,0}$\wedge$}%
}}}}
\put(7501,-10261){\makebox(0,0)[lb]{\smash{{\SetFigFont{6}{7.2}{\familydefault}{\mddefault}{\updefault}{\color[rgb]{0,0,0}$\wedge$}%
}}}}
\put(1501,-1561){\makebox(0,0)[lb]{\smash{{\SetFigFont{6}{7.2}{\familydefault}{\mddefault}{\updefault}{\color[rgb]{0,0,0}$L'_{R}$}%
}}}}
\put(1501,-2161){\makebox(0,0)[lb]{\smash{{\SetFigFont{6}{7.2}{\familydefault}{\mddefault}{\updefault}{\color[rgb]{0,0,0}$U'_{L}$}%
}}}}
\put(8101,-1561){\makebox(0,0)[lb]{\smash{{\SetFigFont{6}{7.2}{\familydefault}{\mddefault}{\updefault}{\color[rgb]{0,0,0}$L'_{L}$}%
}}}}
\put(8101,-2161){\makebox(0,0)[lb]{\smash{{\SetFigFont{6}{7.2}{\familydefault}{\mddefault}{\updefault}{\color[rgb]{0,0,0}$U'_{R}$}%
}}}}
\put(6301,-3961){\makebox(0,0)[lb]{\smash{{\SetFigFont{6}{7.2}{\familydefault}{\mddefault}{\updefault}{\color[rgb]{0,0,0}$L'_{L}$}%
}}}}
\put(6301,-4561){\makebox(0,0)[lb]{\smash{{\SetFigFont{6}{7.2}{\familydefault}{\mddefault}{\updefault}{\color[rgb]{0,0,0}$U'_{R}$}%
}}}}
\put(3301,-5161){\makebox(0,0)[lb]{\smash{{\SetFigFont{6}{7.2}{\familydefault}{\mddefault}{\updefault}{\color[rgb]{0,0,0}$L'_{R}$}%
}}}}
\put(3301,-5761){\makebox(0,0)[lb]{\smash{{\SetFigFont{6}{7.2}{\familydefault}{\mddefault}{\updefault}{\color[rgb]{0,0,0}$U'_{L}$}%
}}}}
\put(6601,-10261){\makebox(0,0)[lb]{\smash{{\SetFigFont{6}{7.2}{\familydefault}{\mddefault}{\updefault}{\color[rgb]{0,0,0}$L'_{L}$}%
}}}}
\put(6601,-10861){\makebox(0,0)[lb]{\smash{{\SetFigFont{6}{7.2}{\familydefault}{\mddefault}{\updefault}{\color[rgb]{0,0,0}$U'_{R}$}%
}}}}
\put(901,-10261){\makebox(0,0)[lb]{\smash{{\SetFigFont{6}{7.2}{\familydefault}{\mddefault}{\updefault}{\color[rgb]{0,0,0}$L'_{R}$}%
}}}}
\put(901,-10861){\makebox(0,0)[lb]{\smash{{\SetFigFont{6}{7.2}{\familydefault}{\mddefault}{\updefault}{\color[rgb]{0,0,0}$U'_{L}$}%
}}}}
\put(7051,-4561){\makebox(0,0)[lb]{\smash{{\SetFigFont{6}{7.2}{\familydefault}{\mddefault}{\updefault}{\color[rgb]{0,0,0}$\wedge$}%
}}}}
\put(7501,-10861){\makebox(0,0)[lb]{\smash{{\SetFigFont{6}{7.2}{\familydefault}{\mddefault}{\updefault}{\color[rgb]{0,0,0}$\wedge$}%
}}}}
\put(4051,-5161){\makebox(0,0)[lb]{\smash{{\SetFigFont{6}{7.2}{\familydefault}{\mddefault}{\updefault}{\color[rgb]{0,0,0}$\wedge$}%
}}}}
\put(4051,-5761){\makebox(0,0)[lb]{\smash{{\SetFigFont{6}{7.2}{\familydefault}{\mddefault}{\updefault}{\color[rgb]{0,0,0}$\wedge$}%
}}}}
\end{picture}%
    
    \caption{Reachable slopes when a weakly exterior point is added to
      a digital straight segment of slope $\frac{1}{2}$. Nodes with an
      odd depth are circled, while those with even depth are
      dash-boxed. $L'_{L}$ and $U'_{L}$ (resp. $L'_{R}$ and $U'_{R}$)
      stand for lower and upper weakly exterior point added to the
      left (resp. right). $\delta$ and $\delta'$ count the number of
      patterns and reversed-patterns
      respectively. \label{fig:reachableNodes}}
    \end{center}
\end{figure}

From \refFigure{fig:evolOddEvenSlopesStern} we can see that the
left child nodes are always reached when we add to the right a lower
weakly exterior point or to the left an upper weakly exterior point
whatever the parity of the node depth. Same reasoning applies for
right child nodes. We can now see how the slope evolution is
translated as moves on the Stern-Brocot tree when adding a point.
\refFigure{fig:reachableNodes} pictures the possible slope evolutions
from the $\frac{1}{2}$ node.

\section{Application to maximal segments}
\label{sec:ms}

We now apply previous properties to get a better understanding of the
geometry of maximal segments. Maximal segments form a particular class
of digital straight segments on the digital curve. Their study is
related to many discrete geometry problems such as digital convexity
\cite{Doerksen04}, polygonalization \cite{Feschet03} or tangent
computation \cite{Lachaud05a}. The main result of this section is
\refTheorem{thm:slopeMinMax} which bounds the slope difference between
two consecutive maximal segments. Let us first explain how to
characterize them.


Given a point $\PT{i}$ on a digital curve $\DC$, the first index $j$
greater than $i$ such that $\SPRED{i}{j}$ and $\neg \SPRED{i}{j+1}$ is
called the {\em front} of $i$. The map associating any $i$ to its
front is denoted by $F$. Symmetrically, the first index $i$ such that
$\SPRED{i}{j}$ and $\neg \SPRED{i-1}{j}$ is called the {\em back} of
$j$ and the associated mapping is denoted by $B$.
\begin{definition}
\label{def:maximal-segment}
Any set of points $\PTS{i}{j}$ is called a {\em maximal segment} iff
any of the following equivalent characterizations holds: (1)
$\SPRED{i}{j}$ and $\neg \SPRED{i}{j+1}$ and $\neg \SPRED{i-1}{j}$, (2)
$B(j)=i$ and $F(i)=j$.
\end{definition}
Maximal segments form the set of DSS on the digital curve that cannot
be extended on any side. They can be ordered along the curve.
Consecutive maximal segments overlap and often on more than two
points. The digital path that belongs to two consecutive maximal
segments is called a \emph{common part}, its associated maximal
segments are $\PTS{B(j)}{j}$ and $\PTS{i}{F(i)}$ if $\PTS{i}{j}$ is
the common part. Note that $\PTS{i-1}{j+1}$ is not a DSS. A common
part is never empty (though it may be reduced to two points). In fact
we know the type of the limiting points for all common parts, as shown
below:
\begin{lemma}  \label{lemma:bothUorL}
  If $\PTS{i}{j}$ is the common part of two consecutive maximal
  segments, then the points $i-1$ and $j+1$ are both upper or both
  lower weakly exterior.
\end{lemma}
\begin{proof}
  $\PTS{i}{F(i)}$ is one of the two consecutive maximal segments, thus
  $F(i) > j$ and $\SPRED{i}{j+1}$ holds. The point $j+1$ is thus an
  interior point or a weakly exterior point for $\PTS{i}{j}$. Assuming
  that $j+1$ is an interior point for $\PTS{i}{j}$, any extensions to
  the back of $\PTS{i}{j}$ is compatible with $\PTS{i}{j+1}$. The
  maximal segment $\PTS{B(j)}{j}$ is one of these extensions, thus
  $\SPRED{B(j)}{j+1}$ would hold which raises a contradiction. As a
  consequence $j+1$ is a weakly exterior point for the DSS
  $\PTS{i}{j}$ and a similar reasoning can be applied to $i-1$.

  
  We prove by contradiction that $i-1$ and $j+1$ are either both lower
  or both upper weakly exterior. Assume $i-1$ is upper weakly exterior
  and $j+1$ is lower weakly exterior. Let the DSS $\PTS{i}{j}$ be
  constituted of $\delta$ patterns and $\delta'$ reversed-patterns. By
  definition, $\delta$ and $\delta'$ differ at most of one but here,
  given the type of $i-1$ and $j+1$, the equality $\delta = \delta'$
  holds.  From \refProposition{prop:dr95}, DSS $\PTS{i-1}{j}$ and
  $\PTS{i}{j+1}$ have the same slope (whichever the parity).
  Furthermore the {\bf DR95} algorithm for updating slopes indicates
  they share the same leaning points $L_{L}$ and $U_{R}$. These two
  assertions, combined together, entail $\PTS{i-1}{j+1}$ is a DSS too,
  which contradicts the hypothesis that it is a common part.  \qed
\end{proof}

We give now analytic bounds on slopes of two consecutive maximal
segments.
\begin{theorem} \label{thm:slopeMinMax} 
  If $\PTS{i}{j}$ is the common part of two consecutive maximal
  segments (namely $\PTS{B(j)}{j}$ and $\PTS{i}{F(i)}$), their slopes
  are such that:
  \begin{center}
    \begin{tabular}{|c|c|c|c|c|c|}
      \cline{3-6}
      \multicolumn{1}{c}{} & \multicolumn{1}{c}{} & \multicolumn{4}{|c|}{$i-1$ and $j+1$ are both (\refLemma{lemma:bothUorL})} \\
      \cline{3-6}
      \multicolumn{1}{c}{} & \multicolumn{1}{c}{} & \multicolumn{2}{|c|}{lower weakly exterior} &  \multicolumn{2}{c|}{upper weakly exterior}\\
      \cline{3-6}
      \multicolumn{1}{c}{} & \multicolumn{1}{c}{} & \multicolumn{1}{|c}{minimal slope}  & \multicolumn{1}{|c}{maximal slope} & \multicolumn{1}{|c}{minimal slope}  & \multicolumn{1}{|c|}{maximal slope} \\
      \cline{1-6}
      \multirow{2}{1.9cm}{$\PTS{i}{j}$ has an even slope} &  
      $\PTS{i}{F(i)}$ & 
      $\frac{ \delta' p_{2i} - p_{2i -1}} { \delta' q_{2i} - q_{2i-1}} $ & 
      $\frac{ ( \delta'+ 2) p_{2i} - p_{2i -1}} { (\delta' +2)q_{2i} - q_{2i-1}}$  & 
      $\frac{ (\delta + 1) p_{2i} + p_{2i -1}} { ( \delta +1 ) q_{2i} + q_{2i-1}}$ & 
      $\frac{ ( \delta -1) p_{2i} + p_{2i -1}} { (\delta -1) q_{2i} + q_{2i-1}}$ \\
      
      \cline{2-6}
      & $\PTS{B(j)}{j}$ & 
      $\frac{ (\delta' + 1) p_{2i} + p_{2i -1}} { ( \delta' +1 ) q_{2i} + q_{2i-1}}$ & 
      $\frac{ ( \delta' -1) p_{2i} + p_{2i -1}} { (\delta' -1) q_{2i} + q_{2i-1}}$ &
      $\frac{ \delta p_{2i} - p_{2i -1}} { \delta q_{2i} - q_{2i-1}} $ & 
      $\frac{ ( \delta+ 2) p_{2i} - p_{2i -1}} { (\delta +2)q_{2i} - q_{2i-1}}$ \\
      
      \cline{1-6}
      \multirow{2}{1.9cm}{$\PTS{i}{j}$ has an odd slope} &  
      $\PTS{i}{F(i)}$ & 
      $\frac{ (\delta' + 1) p_{2i+1} + p_{2i }} { ( \delta' +1 ) q_{2i+1} + q_{2i}}$ & 
      $\frac{ ( \delta' -1) p_{2i+1} + p_{2i}} { (\delta' -1) q_{2i+1} + q_{2i}}$ & 
      $\frac{ \delta p_{2i+1} - p_{2i}} { \delta q_{2i+1} - q_{2i}} $ & 
      $\frac{ ( \delta + 2) p_{2i+1} - p_{2i}} { (\delta +2)q_{2i+1} - q_{2i}}$ \\
      
      \cline{2-6}
      & $\PTS{B(j)}{j}$ & $\frac{ \delta' p_{2i+1} - p_{2i}} { \delta' q_{2i+1} - q_{2i}} $ & 
      $\frac{ ( \delta' + 2) p_{2i+1} - p_{2i}} { (\delta' +2)q_{2i+1} - q_{2i}}$ &
      $\frac{ (\delta + 1) p_{2i+1} + p_{2i }} { ( \delta +1 ) q_{2i+1} + q_{2i}}$ & 
      $\frac{ ( \delta -1) p_{2i+1} + p_{2i}} { (\delta -1) q_{2i+1} + q_{2i}}$ \\
      \hline
    \end{tabular}
  \end{center}
\end{theorem}
\begin{proof}
  The following proof holds if $\PTS{i}{j}$ has an even slope and both
  $i-1$ and $j+1$ are lower weakly exterior points. Other cases are
  deduced from \refProposition{prop:dr95}.

    
  We bound the slopes obtained by extending $\PTS{i}{j}$ to the right
  then extending $\PTS{i}{j}$ to the left. Since $j+1$ is a lower
  weakly exterior point, $\PTS{i}{j+1}$ has slope $z^{R}_{2i+2} =
  [0,u_{1},\ldots,u_{2i}-1,1,\delta']$ (\refProposition{prop:dr95}).
  Assuming that $\PTS{i}{F(i)}$ has a slope that equals
  $[0,u_{1},\ldots,u_{2i}-1,1,\delta'+\epsilon, u_{2i+3}, \ldots ,
  u_{p}]$ with $\epsilon$ being $-1$ or zero (from
  \refProposition{prop:dr95}). Simple calculation brings: \\
  \[ \textstyle{
    \cfrac{1}{u_{2i+3}+\cfrac{1}{\ldots + \cfrac{1}{u_{p}}}} = \epsilon' \quad \quad \textrm{with}~ \epsilon' \in ]0,1] }
  \]

  As a result, the slope of $\PTS{i}{F(i)}$ equals
  $z^{R}(\epsilon_{R})=[0,u_{1},\ldots,u_{2i}-1,1,\delta' +
  \epsilon_{R}]$ with $\epsilon_{R} \in ]-1,1]$. \Equ{pattern:rec:num}
  and \Equ{pattern:rec:den} still hold when partial coefficient are
  real values, we thus get
  \[ \textstyle{
  z^{R}(\epsilon_{R}) = \frac{ ( \delta'+ \epsilon_{R} +1) p_{2i} - p_{2i -1}} { (\delta' + \epsilon_{R}+1)q_{2i} - q_{2i-1}}}
  \]
  We bound this slope for extremal values of $\epsilon_{R}$, giving
  \[\textstyle{
  \frac{ \delta' p_{2i} - p_{2i -1}} { \delta' q_{2i} - q_{2i-1}} \leq z^{R}(\epsilon_{R}) \leq \frac{ ( \delta'+ 2) p_{2i} - p_{2i -1}} { (\delta' +2)q_{2i} - q_{2i-1}} }
  \]
  Same reasoning applied to the back/left of the common part brings:
  $z^{L}_{2i+2} = [0,u_{1},\ldots, u_{2i},\delta']$ and
  $z^{L}(\epsilon_{L})=[0,u_{1},\ldots,u_{2i},\delta' + \epsilon_L]$
  with $\epsilon_L \in ]-1,1]$. Bounds are:
  \[\textstyle{
  \frac{ (\delta' + 1) p_{2i} + p_{2i -1}} { ( \delta' +1 ) q_{2i} + q_{2i-1}} \leq z^{L}(\epsilon_{L}) \leq \frac{ ( \delta' -1) p_{2i} + p_{2i -1}} { (\delta' -1) q_{2i} + q_{2i-1}}.   \qed }
  \]
\end{proof}
Furthermore, \refTheorem{thm:slopeMinMax} give bounds on the slope
difference $\Delta z$ of two consecutive maximal segments as a
function of the parameters of their common part. For instance the case
of an even slope with lower weakly exterior points give the tight bound
\[
\frac{2 \delta' + 3}{(\delta'^{2} +3\delta' + 2)q_{2i}^{2} +
q_{2i}q_{2i-1} - q_{2i-1}^2 } < | \Delta z | < \frac{2 \delta'
-1}{(\delta'^{2} - \delta' )q_{2i}^{2} + q_{2i}q_{2i-1} - q_{2i-1}^2}.
\] 
We give below a coarser bound for $\Delta z$, but which is expressed
only in terms of the slope denominator and the number of
reversed-patterns:
\begin{equation}
\frac{2 \delta' + 3}{(\delta'^{2} +3\delta' + 3)q_{2i}^{2} }
< | \Delta z |  <
\frac{2 \delta' -1}{(\delta'^{2} - \delta')q_{2i}^{2} + \frac{1}{2}q_{2i} + \frac{1}{2}}.
\end{equation}


In the other cases, similar formula are obtained. These formulas
induce that the average slope difference between consecutive maximal
segments could be determined, provided the average behaviour of
$\delta$, $\delta'$ and $q_{n}$ is known.

\section{Conclusion}
\label{sec:conclusion}

We have revisited a classical arithmetically-based DSS recognition
algorithm with new parameters related to a combinatoric representation
of DSS. New analytic relations have been established and the relation
with the Stern-Brocot tree has been made explicit. At last, we have
shown new geometric relations on maximal segments.  The new parameters
introduced in this paper seem to be good candidates to describe DSS
and obtain new properties. It would be interesting to investigate the
average asymptotic behavior of these parameters, that is $\delta$,
$\delta'$ and $q_{n}$, as functional of the grid step. This would lead
us to estimate the asymptotic angle difference between consecutive
maximal segments, a quantity related to curvature, and therefore to
address the problem of finding a multigrid convergent curvature
estimator. This study is thus a first step in this direction. 

\bibliographystyle{plain}

\bibliography{dgci} 



\end{document}